\documentclass[twocolumn,showpacs,floatfix,superscriptaddress,amsmath,amssymb,prl]{revtex4}
\usepackage{amsmath}
\usepackage{amsfonts}
\usepackage{bbm}
\usepackage{mathrsfs}
\usepackage{txfonts}
\usepackage{amssymb}
\usepackage{graphicx}
\usepackage{hyperref}
\usepackage{ulem}
 \usepackage{overpic}
 \usepackage{psfrag}
\usepackage{tabularx}
\usepackage{array}
\usepackage{placeins}
\makeatletter
\renewcommand\subsection{\@startsection
{subsection}{2}{0mm}
 {-\baselineskip}
 {0.5\baselineskip}
{\FloatBarrier\normalfont\Large\bfseries}}
\makeatother
\newcommand{\be}{\begin{equation}}
\newcommand{\ee}{\end{equation}}

\newcommand{\PreserveBackslash}[1]{\let\temp=\\#1\let\\=\temp}

\begin{document}
\title{Ground-state fidelity and quantum criticality in a two-leg ladder with cyclic four-spin exchange}

\author{Sheng-Hao Li} \affiliation{Centre for Modern Physics and Department of Physics,
Chongqing University, Chongqing 400044, The People's Republic of
China}

\author{Qian-Qian Shi} \affiliation{Centre for Modern Physics and Department of Physics,
Chongqing University, Chongqing 400044, The People's Republic of
China}

\author{Jin-Hua Liu} \affiliation{Centre for Modern Physics and Department of
Physics, Chongqing University, Chongqing 400044, The People's
Republic of China}

\author{Huan-Qiang Zhou}\affiliation{Centre for Modern Physics and Department of Physics,
Chongqing University, Chongqing 400044, The People's Republic of
China}

\begin{abstract}
We investigate a two-leg Heisenberg spin ladder with cyclic
four-spin exchange by exploiting a newly-developed tensor network
algorithm. The algorithm allows to efficiently compute the
ground-state fidelity per lattice site, which enables us to
establish the ground-state phase diagram for quantum lattice
many-body systems. The latter is based on the observation that, for
an infinite-size system, any singularity on a ground-state fidelity
surface characterizes a critical point, at which the system
undergoes a phase transition. For the two-leg Heisenberg spin-$1/2$
ladder with cyclic four-spin exchange, six different phases are
identified: the ferromagnetic phase, the rung singlet phase, the
staggered dimer phase, the scalar chirality phase, the dominant
vector chirality region, and the dominant collinear spin region. Our
findings are in a good agreement with the previous studies from the
exact diagonalization and the density-matrix renormalization group.
\end{abstract}
 \maketitle
%%%%%%%%%%%%%%%%%%%%%%%%%%%%%%%%%%%%%%%%%%%%%%
{\it Introduction.} In the last decades, low-dimensional quantum
spin systems, such as spin ladders, have been the subject of
extensive experimental and theoretical interest. Many fascinating
features of the ladder systems have long been understood
theoretically from both analytical and numerical
approaches~\cite{giamarchi,dagotto}. Among them is an intriguing
property that the existence of an excitation gap depends on the
number of legs: spin excitations are gapful for an even-leg ladder,
and gapless for an odd-leg ladder. There are also a number of
low-dimensional cuprate compounds of transition metals, whose
properties are described adequately by multi-leg spin ladders. These
ladder systems often exhibit some attractive properties such as
quantum criticality. In addition, properties of several classes of
materials such as $SrCu_{2}O_{3}$~\cite{a8},
$La_{6}Ca_{8}Cu_{24}O_{41}$~\cite{a7}, and
$(C_{5}H_{12}N)_{2}CuBr_{4}$, are well described by a two-leg
Heisenberg ladder with cyclic four-spin exchange.

On the other hand, a novel approach to critical phenomena in quantum
many-body physics has emerged, based on the fact that fidelity, a
basic notion in quantum information science, is a measure of quantum
state distinguishability. The approach allows us to characterize
critical phenomena in a variety of quantum many-body lattice systems
in any spatial
dimensions~\cite{b0,b1,b2,b3,b4,b5,whl,fidelity1,you,rams}. As
argued in Refs.~\cite{b1,b2,b3,b4,b5,whl}, the ground-state fidelity
per lattice site is able to capture drastic changes of the
ground-state wave functions around a critical point. This, in
combination with the fact that many powerful numerical algorithms
have been developed in the context of the tensor network (TN)
representation, provides a powerful means to unveil quantum
criticality underlying quantum many-body systems. In fact, a
systematic scheme to study critical phenomena in quantum many-body
lattice systems consists of three steps, as advocated in
Ref.~\cite{zhou-op}: first, map out the ground-state phase diagram
by computing the ground-state fidelity per lattice site; second,
derive local order parameters (if any) from the reduced density
matrices for a representative ground-state wave function in a given
phase; third, characterize any phase without any long range order.

In this paper, we consider a two-leg Heisenberg spin-$1/2$ ladder
with cyclic four-spin exchange. The ladder system exhibits a very
rich phase diagram, with six different phases identified: the
ferromagnetic phase, the rung singlet phase, the staggered dimer
phase, the scalar chirality phase, the dominant vector chirality
region, and the dominant collinear spin region, when a properly
chosen control parameter is varied. Therefore, the ladder system
provides a test bed for our scheme to study quantum criticality in
spin ladders. This is achieved by exploiting a newly-developed TN
algorithm~\cite{shli}, which allows to efficiently compute the
ground-state fidelity per lattice site, and the reduced density
matrix from a representative ground-state wave function. Our
findings are in a good agreement with the previous studies from the
exact diagonalization~\cite{a1,a4} and the density-matrix
renormalization group (DMRG)~\cite{a4}.

{\it Tensor network representation for spin ladders.} The TN
representation is a convenient way to represent ground-state wave
functions in classical simulations of quantum many-body lattice
systems, such as the matrix product state (MPS)~\cite{b6,b7,
b8,TEBD,iTEBD} in one spatial dimension and the projected
entangled-pair state (PEPS)~\cite{PEPS,iPEPS,b10,b11} in two and
higher spatial dimensions. Here, let us briefly recall the gradient
TN algorithm to compute the ground-state wave functions for quantum
many-body systems on an infinite-size  two-leg spin
ladder~\cite{shli}, adapted to be suitable to a system with cyclic
four-spin exchange.

Assume that the Hamiltonian is translation-invariant:
$H=\sum_{i}h^{[i]}$, with $h^{[i]}$ being the $i$-th plaquette
Hamiltonian density along the leg direction. A TN representation for
a quantum wave function consists of  four-index tensors $A_{\ell
rd}^{s}$, $B_{\ell rd}^{s}$, $C_{\ell ru}^{s}$, and $D_{\ell
ru}^{s}$ attached to each site in the unit cell. Here, $s$ is a
physical index, $s=1,...,\mathbbm{d}$, with $\mathbbm{d}$ being the
dimension of the local Hilbert space, and $\ell$, $r$, $u$, and $d$
denote bond indices, $\ell$, $r$, $u$, $d=1,...,\mathbb{D}$, with
$\mathbb{D}$ being the bond dimension. Then, for a random initial
state $|\psi_0\rangle$, the energy is a functional of the TN
tensors:
\begin{equation}
E=\frac{\langle \psi_0 | H |\psi_0 \rangle}{\langle \psi_0 |\psi_0
\rangle},
\end{equation}
which allows an efficient computation in the context of the TN
representation. To update four-index tensors $A_{\ell rd}^{s}$,
$B_{\ell rd}^{s}$, $C_{\ell ru}^{s}$, and $D_{\ell ru}^{s}$, we need
to compute the energy gradient,
\begin{align}
\frac{\partial E}{\partial X_{\ell rd}^{s}}=\frac {1}{\langle \psi_0
|\psi_0 \rangle} \frac{{\partial \langle \psi_0 | H |\psi_0
\rangle}}{{\partial X_{\ell rd}^{s}}}- \frac {E}{\langle \psi_0
|\psi_0 \rangle} \frac{{\partial \langle \psi_0 |\psi_0
\rangle}}{{\partial X_{\ell rd}^{s}}},
\end{align}
where $X_{\ell rd}^{s} \in \{A_{\ell rd}^{s}, B_{\ell rd}^{s},
C_{\ell ru}^{s}, D_{\ell ru}^{s} \}$. Once the energy gradient is
known, the four-index tensors $A_{\ell rd}^{s}$, $B_{\ell rd}^{s}$,
$C_{\ell ru}^{s}$, and $D_{\ell ru}^{s}$ may be updated as follows,
\begin{align}
X_{\ell rd}^{s}=X_{\ell rd}^{s} - \delta \frac{\partial E}{\partial
X_{\ell rd}^{s}}.
\end{align}
Here, $\delta$ denotes the step size during each iteration, which is
tuned to be decreasing in the implementation, when the ground-state
wave function is approached. Here, we stress that four different
tensors $A_{\ell rd}^{s}$, $B_{\ell rd}^{s}$, $C_{\ell ru}^{s}$, and
$D_{\ell ru}^{s}$ are updated simultaneously. Repeating this
updating procedure until the ground-state energy converges, the
system's ground-state wave function is generated in the TN
representation. Actually, if the energy gradient $\frac{\partial
E}{\partial X}\rightarrow 0$, a good approximation to the
ground-state wave function is anticipated.

{\it The model.} The two-leg Heisenberg spin-$1/2$ ladder with
cyclic four-spin exchange is described by the Hamiltonian:
\begin{equation}\label {fitlader1}
\begin{split}
   H=&J_{\bot}\sum_{i} S_{1, i}\cdot S_{2, i}+J_{\|}\sum_{i} (S_{1, i}\cdot S_{1,
   i+1}+S_{2, i}\cdot S_{2,i+1})\\
   &+K\sum_{i} (P_{i, i+1}+P^{-1}_{i, i+1}).
    \end{split}
 \end{equation}
Here, $S_{\alpha,i}$ ($\alpha=1,2$) denotes the spin-$1/2$ Pauli
operators at site $i$ on the $\alpha$-th leg, $J_{\bot}$ is the
interchain coupling between two spins on each rung, $J_{\|}$ is the
intrachain coupling between two neighboring spins in each chain, and
$K$ is the cyclic four-spin exchange interaction coupling. The
cyclic four-spin permutation operator $P_{i, i+1}$ ($P^{-1}_{i,
i+1}$) exchanges the four spins around the $i$-th plaquette as
$S_{1, i}\rightarrow S_{1, i+1}\rightarrow S_{2, i+1}\rightarrow
S_{2, i}\rightarrow S_{1, i}$, which can be decomposed in terms of
the Pauli spin operators involving bilinear and biquadratic terms:
\begin{equation}\label {fitlader2}
\begin{split}
   P_{i, i+1}+P^{-1}_{i, i+1}=&S_{1, i}\cdot S_{1, i+1}+S_{1, i+1}\cdot S_{2, i+1}+S_{2, i+1}\cdot S_{2,
   i}\\
   &+S_{2, i}\cdot S_{1,i}+S_{1, i}\cdot S_{2, i+1}+S_{1, i+1}\cdot S_{2, i}\\
   &+4(S_{1, i}\cdot S_{1, i+1})(S_{2, i+1}\cdot S_{2, i})\\
   &+4(S_{2, i}\cdot S_{1,i})(S_{1, i+1}\cdot S_{2, i+1})\\
   &-4(S_{1, i}\cdot S_{2, i+1})(S_{1,i+1}\cdot S_{2, i}).
    \end{split}
 \end{equation}
The model has been investigated by the exact
diagonalization~\cite{a1,a11} and DMRG~\cite{a1,a12,a121,a13,a4}.
Here, we focus on the computation of the ground-state fidelity per
lattice site in terms of the newly-developed TN algorithm. For
simplicity, we choose $K=\sin\theta$, and
$J_{\bot}=J_{\|}=\cos\theta$, with $\theta\in[-\pi,\pi]$.

{\it The ground-state fidelity per lattice site.} The ground-state
wave function is generated from the newly-developed TN algorithm for
a given choice of the coupling constants of the spin ladder, which
allows to efficiently evaluate the ground-state fidelity per lattice
site, a universal marker to detect quantum criticalities: a phase
transition point is characterized by a pinch point on the fidelity
surface.

For the two-leg Heisenberg spin ladder with cyclic four-spin
exchange, we choose $\theta$ as a control parameter. For two
different ground states, $|\psi(\theta_1)\rangle$ and
$|\psi(\theta_2)\rangle$, corresponding to two different values
$\theta_1$ and $\theta_2$ of the control parameter $\theta$, the
ground-state fidelity
$F(\theta_1,\theta_2)=|\langle\psi(\theta_2)|\psi(\theta_1)\rangle|$
asymptotically scales as $F(\theta_1,\theta_2)\sim
d(\theta_1,\theta_2)^N$, with $N$ the total number of the lattice
sites. Here, $d(\theta_1,\theta_2)$ is the scaling parameter,
introduced in Refs.\cite{b1,b2,b3} for one-dimensional quantum
lattice systems and in Refs.\cite{b4} for two and higher-dimensional
quantum lattice systems. In fact, $d(\theta_1,\theta_2)$
characterizes how fast the fidelity goes to zero when the
thermodynamic limit is approached. Physically, the scaling parameter
$d(\theta_1,\theta_2)$ is the averaged fidelity per lattice site,
\begin{equation}
\ln d(\theta_1,\theta_2)\equiv \lim_{N\rightarrow \infty}\frac{\ln
F(\theta_1,\theta_2)}{N},
\end{equation}
which is well defined in the thermodynamic limit even if
$F(\theta_1,\theta_2)$ becomes trivially zero. It satisfies the
properties inherited from the fidelity $F(\theta_1,\theta_2)$: (i)
normalization $d(\theta,\theta)= 1$; (ii) symmetry
$d(\theta_1,\theta_2)= d(\theta_2,\theta_1)$; and (iii) range $0
\leq d(\theta_1,\theta_2) \leq 1$.

%%%%%%%%%%%%%%%%%%%%%%%%%%%%%%%%%%%%%%%%%%%%%%%%%%%%%%%%%%%%%%%%%%%%%%
\begin{figure}
\begin{center}
\includegraphics[width=0.40\textwidth]{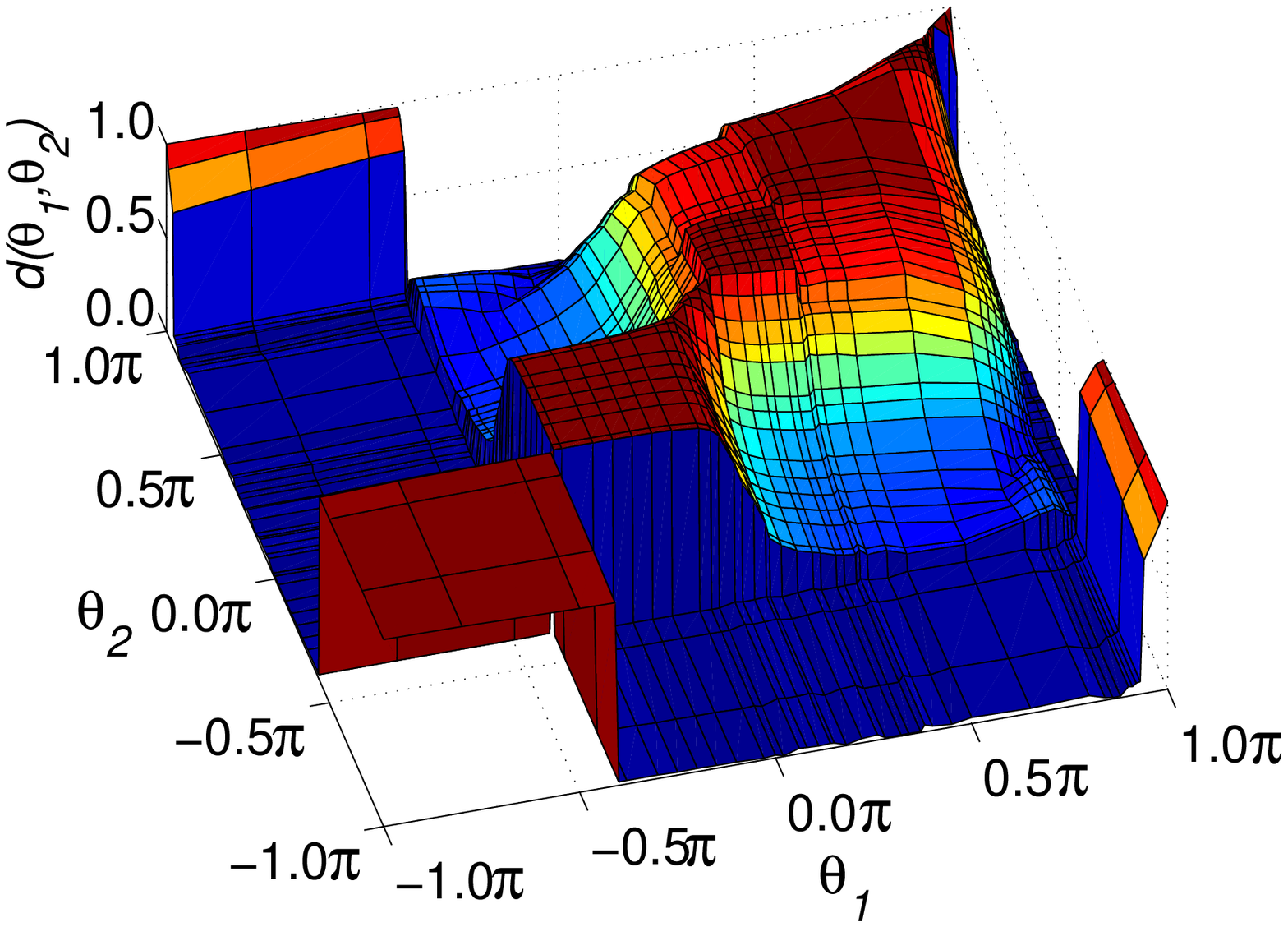}
\end{center}
\vspace*{-0.5cm}
\begin{center}
\includegraphics[width=0.35\textwidth]{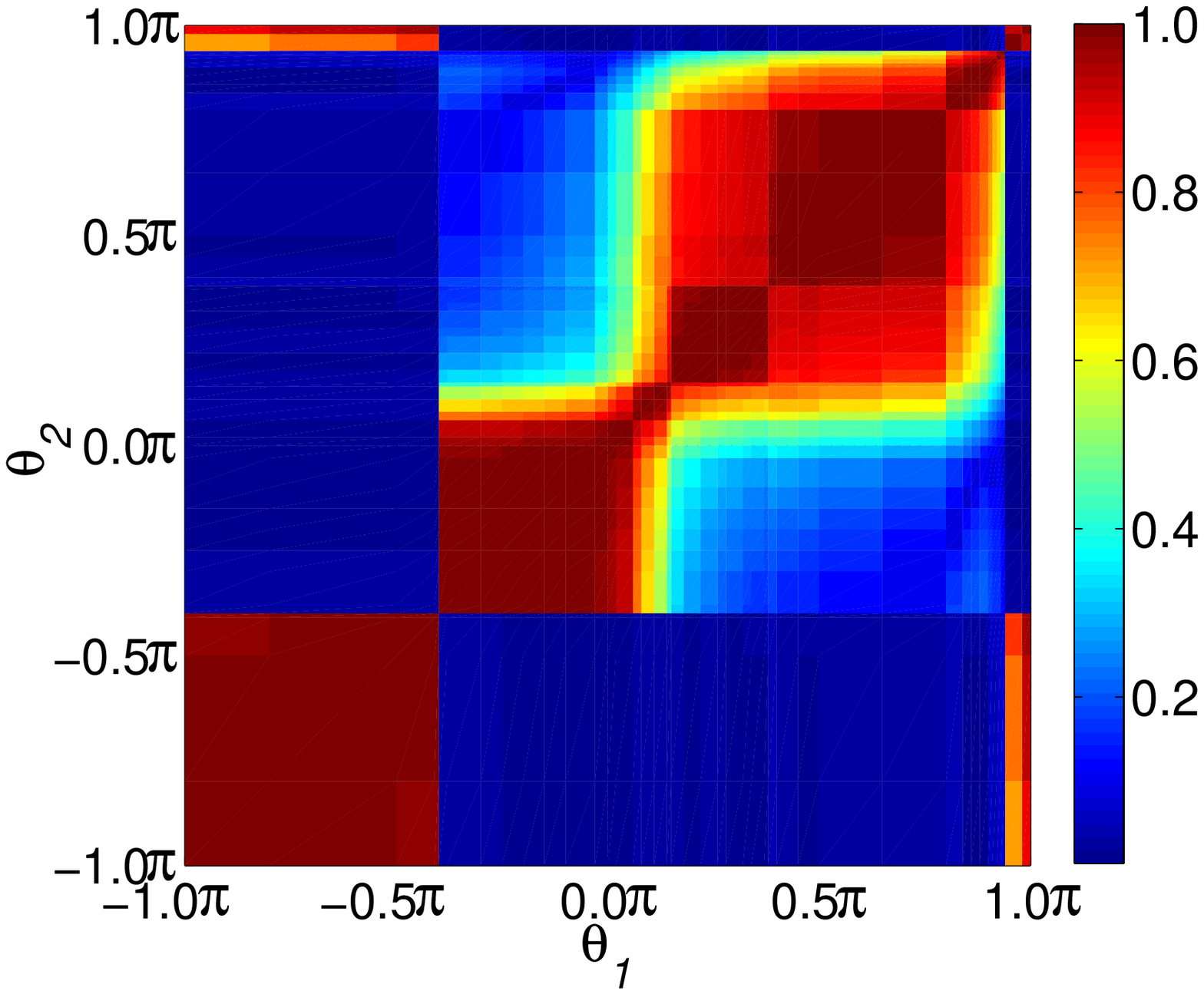}
\end{center}
\vspace*{-0.5cm}
\caption{(color online) Upper panel: A
two-dimensional fidelity surface embedded in a three-dimensional
Euclidean space. Lower panel: The countour plot of the ground-state
fidelity per lattice site, $d(\theta_1, \theta_2)$, on the
($\theta_1, \theta_2$)-plane, for the two-leg Heisenberg spin-$1/2$
ladder with cyclic four-spin exchange. There are six pinch points on
the fidelity surface. Therefore, six different phases are
identified: the ferromagnetic phase, the rung singlet phase, the
staggered dimer phase, the scalar chirality phase, the dominant
vector chirality region, and the dominant collinear spin region. The
ground-state phase diagram is as follows: the ferromagnetic phase
for $-1.06\pi(0.94\pi)\lesssim \theta \lesssim -0.40\pi$, the rung
singlet phase for $-0.40\pi\lesssim \theta \lesssim 0.06\pi$, the
staggered dimer phase for $0.06\pi\lesssim \theta \lesssim 0.15\pi$,
the scalar chirality phase for $0.15\pi\lesssim \theta \lesssim
0.38\pi$, the dominant vector chirality region for $0.38\pi\lesssim
\theta \lesssim 0.80\pi$, and the dominant collinear spin region for
$0.80\pi\lesssim \theta \lesssim 0.94\pi$.} \label{phasediagram}
\end{figure}

In Fig.\ref{phasediagram}, we plot a two-dimensional fidelity
surface embedded in a three-dimensional Euclidean space for the
two-leg Heisenberg spin-$1/2$ ladder with cyclic four-spin exchange.
As shown in the upper panel, there are six pinch points on the
fidelity surface, implying six phase transition points. In the lower
panel, the contour plot of the ground-state fidelity per lattice
site $d(\theta_1, \theta_2)$ on the ($\theta_1, \theta_2$)-plane is
shown, with the truncation dimension up to $4$. Therefore, we are
able, by evaluating the ground-state fidelity per lattice site, to
identify six different phases: the ferromagnetic phase, the rung
singlet phase, the staggered dimer phase, the scalar chirality
phase, the dominant vector chirality region, and the dominant
collinear spin region. Notice that, among six transition points,
there are two first-order phase transitions at
$\theta\approx-0.40\pi$ and $\theta\approx0.94\pi$ between
ferromagnetic phase and its adjacent phases: the rung singlet phase
and the dominant collinear spin region. The remaining four
transition points are continuous. These results are in a good
agreement with the earlier analyses~\cite{a1,a4} based on the exact
diagonalization and DMRG. Therefore, the TN algorithm yields
reliable results for the two-leg Heisenberg spin ladder with cyclic
four-spin exchange. In addition, the ground-state fidelity per
lattice site, as a universal marker to detect phase transitions, is
able to capture drastic changes of ground-state wave functions
around critical points for quantum spin ladders.

{\it Order parameters.} Once the ground-state phase diagram is
known, we are able to read out local order parameters (if any) from
the reduced density matrices for a representative ground-state wave
function in a given phase, as advocated in Ref.~\cite{zhou-op}.

\begin{figure}
\begin{center}
\includegraphics[height=45mm,width=0.40\textwidth]{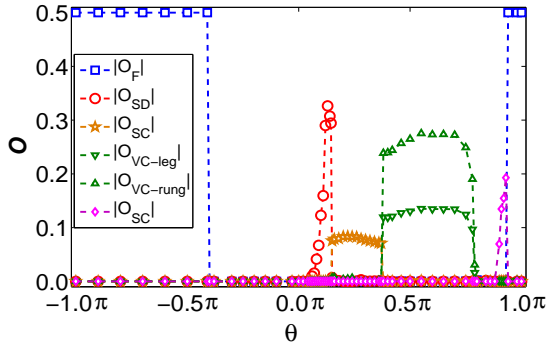}
\end{center}
\vspace*{-0.5cm}
\caption{(color online) The local order parameters
$O_{F}$, $O_{SD}$, $O_{SC}$,  $O_{VC}$, and $O_{CS}$ in the
ferromagnetic phase , the staggered dimer phase, the scalar
chirality phase, the dominant vector chirality region, and the
dominant collinear spin region versus $\theta$, respectively.}
\label{SCSD}
\end{figure}

In the ferromagnetic phase, the non-zero-entry structure of the
one-site reduced density matrix shows that the $\langle S_{\alpha,
i} \rangle$ are the same at all the lattice sites for the two-leg
spin ladder. Therefore, the local order parameter is
\begin{equation}
O_{F}=\langle \psi_0 |S_{\alpha, i}|\psi_0 \rangle.
\end{equation}
As seen in Fig.~\ref{SCSD}),  $|O_{F}|\equiv0.50$. That is, spins
are fully polarized in this phase. In fact, the spin correlations
$\langle S_{\alpha, i}\cdot S_{\alpha, i+1}\rangle$ between the
nearest-neighbor spins on the legs and the spin correlations
$\langle S_{1, i}\cdot S_{2, i}\rangle$ between spins on the rungs
are 0.25. The ferromagnetic state minimizes the energy on each
plaquette separately for $-1.06\pi(0.94\pi)\lesssim \theta \lesssim
-0.40\pi$.

In the rung singlet phase, the ground-state wave function may be
approximated by the product of local rung singlets. The ground-state
lacks long-range order in the conventional sense, thus there is no
local order parameter; instead, an exotic order occurs. In fact, the
rung singlet phase and the Haldane phase are essentially the same,
namely, both are characterized by the so-called string
order~\cite{a5}. The rung singlet phase lies in $-0.40\pi\lesssim
\theta \lesssim 0.06\pi$.

In the staggered dimer phase, the non-zero-entry structure of the
two-site reduced density matrix exhibits a pattern, with the local
order parameter as follows,
\begin{equation}\label {fitlader3}
\begin{split}
O_{SD}=&\frac{1}{2} \langle \psi_0 |S_{1, i-1}\cdot S_{1, i}-S_{1,
i}\cdot S_{1,
i+1}\\
 &+S_{2, i}\cdot S_{2, i+1} -S_{2, i-1}\cdot S_{2, i}|\psi_0 \rangle.
\end{split}
\end{equation}
Here, $\langle \psi_0 |S_{\alpha, i}\cdot S_{\alpha, i+1}| \psi_0
\rangle=\langle \psi_0 |S_{1, 1}\cdot S_{1, 2}| \psi_0\rangle$ if
$\alpha+i$ is even, $\langle \psi_0 |S_{\alpha, i}\cdot S_{\alpha,
i+1}| \psi_0\rangle=\langle \psi_0 |S_{2, 1}\cdot S_{2, 2}|
\psi_0\rangle$ if $\alpha+i$ is odd for two degenerate
symmetry-breaking ground states in this phase. The ladder is in the
staggered dimer phase for $0.06\pi\lesssim \theta \lesssim 0.15\pi$
(cf. Fig.\ref{SCSD}).

In the scalar chirality phase, we need to study the non-zero-entry
structure of the three-site reduced density matrix. This yields the
local order parameter
\begin{equation}
O_{SC}=\langle \psi_0 |S_{1, i}\cdot (S_{2, i}\times S_{1,
i+1})|\psi_0 \rangle.
\end{equation}
It breaks the spatial symmetries and the time reversal symmetry, but
not the internal SU(2) symmetry. The scalar chirality phase lies in
$0.15\pi\lesssim \theta \lesssim 0.38\pi$, as seen from
Fig.\ref{SCSD}.

In the dominant vector chirality region, the non-zero-entry
structure of the two-site reduced density matrix yields the local
order parameter
\begin{equation}
O_{VC-leg}=\langle \psi_0 |S_{\alpha, i}\times S_{\alpha,
i+1}|\psi_0 \rangle,
\end{equation}
\begin{equation}
O_{VC-rung}=\langle \psi_0 |S_{\alpha, i}\times S_{\alpha+1,
i}|\psi_0 \rangle.
\end{equation}
It breaks the spatial symmetries and the time reversal symmetry. In
this phase, the spin correlations are strong between bonds on rungs
and legs, but the spin correlations are very weak between diagonal
bonds. The order parameter is plotted in Fig.\ref{SCSD}, which is
non-zero between $0.38\pi\lesssim \theta \lesssim 0.80\pi$.

In the dominant collinear spin region, spins on the same leg exhibit
ferromagnetic correlations, while spins on the same rung exhibit
antiferromagnetic correlations. The non-zero-entry structure of the
one-site reduced density matrix yields the local order parameter
\begin{equation}
O_{CS}=\frac{1}{2}\langle \psi_0 |S_{1, i}-S_{2, i}|\psi_0 \rangle.
\end{equation}
The dominant collinear spin region lies in $0.80\pi\lesssim \theta
\lesssim 0.94\pi$.
%%%%%%%%%%%%%%%%%%%%%%%%%%%%%%%%%%%%%%%%%%%
%\begin{figure}
%\centering
%\begin{overpic}
% [width=0.40\textwidth,totalheight=45mm]{SDC.eps}
%  \end{overpic}

%%%%%%%%%%%%%%%%%%%%%%%%%%%%%%%%%%%%%%%%%%%%

Therefore, we are able to ``derive'', by investigating the
non-zero-entry structure of the reduced density matrices for
representative ground-state wave functions from different phases,
the local order parameters for the ferromagnetic phase, the
staggered dimer phase, the scalar chirality phase, the dominant
vector chirality region and the dominant collinear spin region, with
the order parameter $O_{F}$, $O_{SD}$, $O_{SC}$, $O_{VC}$ and
$O_{CS}$ explicitly shown in Fig.\ref{SCSD}. In addition, no local
order parameter is detected in the rung singlet phase, indicating
that long range order is lacking in this phase. The ground-state
phase diagram established from the local order parameters coincides
with that from the ground-state fidelity per lattice site. That is,
the ladder system undergoes four continuous phase transitions at
$\theta\approx0.06\pi$, $\theta\approx0.15\pi$,
$\theta\approx0.38\pi$ and $\theta\approx0.80\pi$, and two
first-order phase transitions at $\theta\approx-0.40\pi$ and
$\theta\approx0.94\pi$.

{\it Conclusions.} We have exploited a newly-developed efficient TN
algorithm to compute ground-state wave functions for the two-leg
Heisenberg ladder with cyclic four-spin exchange. We have mapped
out, by computing the ground-state fidelity per lattice site, the
phase diagram of the two-leg spin ladder with cyclic four-spin
exchange. Six different phases are identified: the ferromagnetic
phase, the rung singlet phase, the staggered dimer phase, the scalar
chirality phase, the dominant vector chirality phase, and the
dominant collinear spin phase. In addition, the corresponding local
order parameters  are ``derived" from the reduced density matrices,
which are computed efficiently in the context of the TN
representation. Our findings are in a good agreement with the
previous studies from the exact diagonalization~\cite{a1,a4} and
DMRG~\cite{a4}. Therefore, the TN algorithm yields reliable results,
and the ground-state fidelity per lattice site, as a universal
marker to detect phase transitions, is able to capture drastic
changes of ground-state wave functions around critical points, even
for the two-leg Heisenberg spin ladder with cyclic four-spin
exchange.

{\it Acknowledgments.} We thank Sam Young Cho, Bing-Quan Hu, Bo Li,
Hong-Lei Wang, Ai-Min Chen, Yao-Heng Su, and Jian-Hui Zhao for
enlightening discussions. This work is supported by the National
Natural Science Foundation of China (Grant No: 10874252). SHL, QQS,
and JHL are supported by Chongqing University Postgraduates' Science
and Innovation Fund (Project No.: 200911C1A0060322) and by the
Fundamental Research Funds for the Central Universities (Project No.
CDJXS11102214).

\end{document}